\documentclass[twocolumn,prl,floatfix,nobibnotes,footinbib]{revtex4-1}
\usepackage{hyperref}
\usepackage{amsmath}
\usepackage{bbold}
\usepackage{amssymb}
\usepackage{graphicx}
\usepackage{bm}
\usepackage{color}
\newcommand{\editRR}[1] {\textcolor{black}{#1}}
\newcommand{\edit}[1] {\textcolor{black}{#1}}

\newcommand{\editE}[1] {\textcolor{black}{#1}}
\newcommand{\editEE}[1] {\textcolor{black}{#1}}

\begin{document}

\title{Majorana oscillations and parity crossings in semiconductor-nanowire-based transmon qubits}
\author{J. \' Avila$^1$, E. Prada$^2$, P. San-Jose$^1$, R. Aguado$^1$}
\affiliation{\\
$^1$Instituto de Ciencia de Materiales de Madrid (ICMM), Consejo Superior de Investigaciones Cient\'{i}ficas (CSIC), Sor Juana In\'{e}s de la Cruz 3, 28049 Madrid, Spain. Research Platform on Quantum Technologies (CSIC).\\$^2$Departamento de F\'isica de la Materia Condensada, Condensed Matter Physics Center (IFIMAC) and Instituto Nicol\'as Cabrera, Universidad Aut\'onoma de Madrid, E-28049 Madrid, Spain}

\date{\today}
\begin{abstract}
We show that the microwave (MW) spectra  in semiconductor-nanowire-based transmon qubits provide a strong signature of the presence of Majorana bound states in the junction. This occurs as an external magnetic field tunes the wire into the topological regime and the energy splitting of the emergent Majorana modes oscillates around zero energy owing to their wave function spatial overlap in finite-length wires. In particular, we discuss how these Majorana oscillations, and the concomitant fermion parity switches in the ground state of the junction, result in distinct spectroscopic features --in the form of an intermitent visibility of absorption lines-- that strongly deviate from standard transmon behavior.  In contrast, non-oscillating zero modes, such as topologically trivial Andreev bound states resulting from sufficiently smooth potentials, exhibit an overall standard transmon response. These differences in the MW response could help determine whether the junction contains topological Majoranas or not.
\end{abstract}\maketitle


\emph{Introduction}--Superconducting islands based on Josephson junctions (JJs)  shunted by a capacitor are the key element in qubits based on superconducting circuits \cite{PhysRevA.69.062320,Devoret1169,Wendin_2017}.
Their physics is controlled by the ratio $E_J/E_C$ between the Josephson coupling $E_J$ and the charging energy $E_C$. This interplay is described by the Hamiltonian {$H=4E_C(\hat N-n_g)^2+V_J(\hat\varphi)$, where $V_J(\hat\varphi)=-E_J\cos\hat \varphi$ \cite{Bouchiat_1998}. Here} $\hat N$ is the Cooper pair number, conjugate to the superconducting phase  $\hat\varphi$, and $n_g=Q_g/2e$ is a gate-induced offset.
Recent experimental efforts are pushing the standard operation limits in order to have JJs compatible with electrical gating and high magnetic fields. This compatibility is a crucial step to reach a regime relevant for microwave (MW) readout of topological qubits based on Majorana bound states (MBSs)~\cite{Hassler_2011,PhysRevB.88.235401,PhysRevLett.111.107007,PhysRevB.88.144507,ginossar2014microwave,PhysRevB.92.075143,PhysRevB.92.245432,PhysRevB.92.134508,PhysRevB.94.085409,PhysRevLett.118.126803,PhysRevB.97.041415,10.21468/SciPostPhys.7.4.050}. Various options include semiconductors \cite{PhysRevLett.115.127001,PhysRevLett.115.127002,PhysRevB.97.060508,PhysRevLett.116.150505,PhysRevLett.120.100502,PhysRevB.99.085434,Casparis2018} and van der Waals heterostructures \cite{Kroll2018,Schmidt2018,Wang2019}. 

We here focus on JJs based on semiconducting nanowires (NWs). \editRR{In the presence of an external Zeeman field $B$, the NWs can be driven into a topological superconductor phase \cite{Lutchyn:PRL10,Oreg:PRL10,Aguado:RNC17,Lutchyn:NRM18} where MBSs couple across the JJ with strength $E_M$} and coherently interact with the superconducting island degrees of freedom.\editRR{ to study the \editEE{detailed} magnetic field evolution of the NW junction, including the emergence of MBSs in the topological phase, and to address the possibility of non-topological robust zero modes that appear in junctions with spatially smooth potentials \cite{Moore:PRB18,Penaranda:PRB18,Vuik:SP19,Avila:CP19,prada2019}. Since both $E_J$ and $E_M$ depend on $B$, this allows to unveil new physics  from the $E_J/E_C\lesssim 1$ Cooper pair box (CPB) to the $E_J\gg E_C$ transmon regimes depending on the ratio $E_M/E_C$.
The hitherto unexplored $E_M/E_C\gg 1$ regime presents parity-dependent MW signatures that map parity switches in the NW spectrum owing to the oscillatory energy splitting of overlapping Majoranas. In contrast, non-topological modes \editEE{pinned to zero energy} yield a standard transmon spectrum. Hence, MW spectroscopy provides a powerful tool to distinguish between both scenarios.}

\begin{figure}[h!]
\includegraphics[width=\columnwidth]{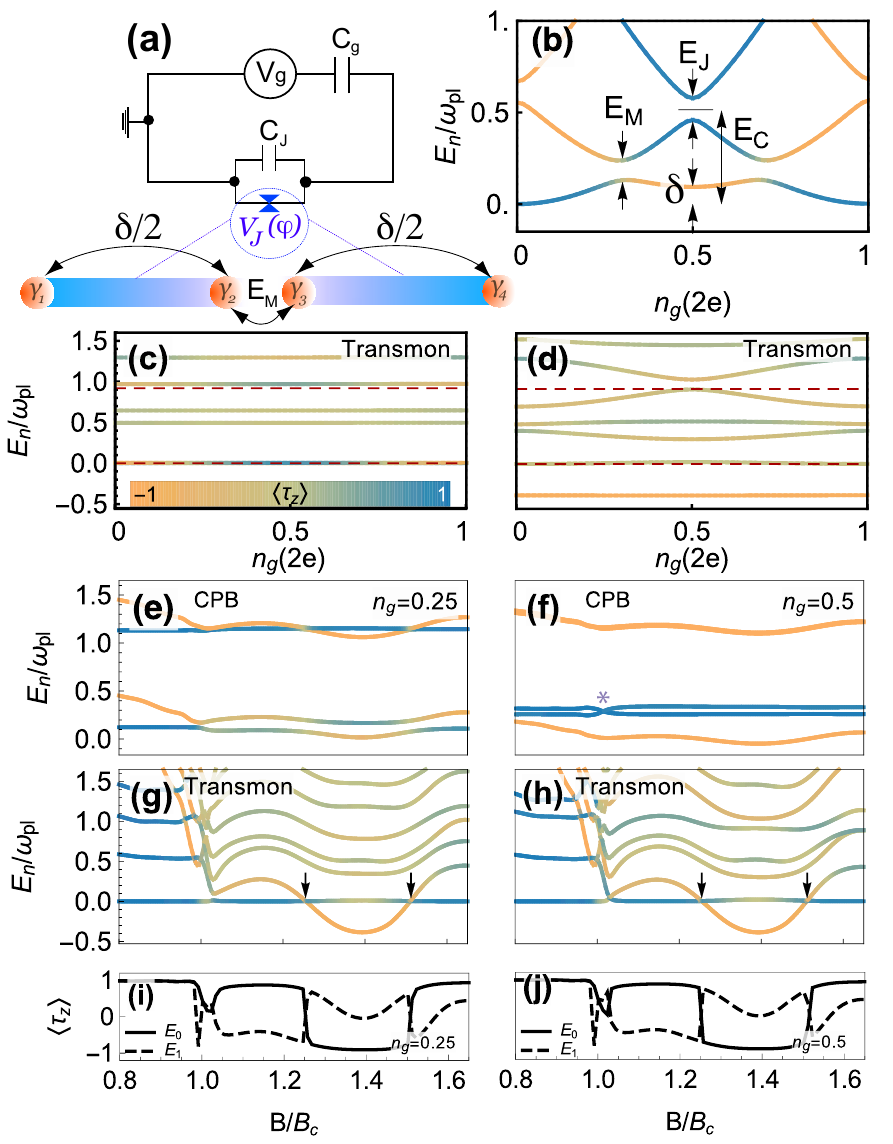}
\caption{\label{fig:evenodd} \textbf{Sketch and spectrum of a NW-based superconducting island.} \editRR{(a) Simplified transmon/CPB circuit. $E_J$ is implicit in the Josephson potential $V_J(\varphi)$, while the combination of a shunting capacitor $C_J$ and the gate capacitance $C_g$ define the charging energy $E_C=e^2/2(C_J+C_g)$. The  presence of MBSs $\gamma_i$ gives rise to two new energy scales, the Majorana splitting $\delta/2$ and the Majorana coupling $E_M$. (b) Spectrum of the island versus $n_g=V_g/(2e C_g)$ in the CPB regime ($E_J/E_C=0.5$, $E_M/E_C\sim 0.12$) and for $B\sim B_c$. }Blue/orange colors denote fermionic even/odd parities, while intermediate gradient signals parity mixing. \editRR{The dotted lines show a slightly different $B$ field with $\delta=0$}. 
(c) Same as (b) but in a transmon regime with $E_J/E_C=25$, $E_M/E_C\sim 6.5$ (dashed lines show the original transmon lines at $B=0$). 
Panel (d) shows the same transmon case but for $\delta\neq 0$ ($B=1.4B_c$). (e,f) Magnetic field dependence for the CPB in (b), at $n_g=0.25$ and $n_g=0.5$, respectively. \editRR{In this latter case, the topological transition (where $E_J\rightarrow 0$) can be seen as a closing of the qubit frequency (asterisk)}. (g,h) Same as (e,f) but for the transmon (arrows mark parity crossings). (i,j) Parity polarization $\langle\hat\tau_z\rangle$ of the two lowest states in (g,h). At parity crossings, the polarization jumps. Rest of parameters: $L_S=2.2\mu$m, $\tau=0.8$, $\mu=0.5$meV, $\Delta=0.25$meV.}
\end{figure}

\emph{Model}--The Josephson potential is defined, {on a microscopic level,} as the {operator} 
$V_J(\varphi)=\frac{1}{2}\check{\bm{c}}^\dagger H_\mathrm{BdG}(\varphi)\check{\bm{c}}$, where $\check{\bm c} = (c_{i\uparrow},c_{i\downarrow},c^\dagger_{i\uparrow},c^\dagger_{i\downarrow})$ are Nambu spinors and $H_\mathrm{BdG}$ is the Bogoliubov-de Gennes (BdG) Hamiltonian
\begin{equation}
\label{microscopic}
H_\mathrm{BdG}(\varphi)=\begin{pmatrix}
H_{\rm{NW}}&\Delta(x,\varphi)\\
\Delta(x,\varphi)^\dagger&-H_{\rm{NW}}^*
\end{pmatrix}.
\end{equation}
$H_{\rm{NW}}$ consists of two (left/right) segments of length $L_S$ with normal Hamiltonians $H_{L/R}$, {coupled across a short weak link of transparency $T_N\in[0,1]$}. Each NW segment contains all the microscopic details (Rashba coupling $\alpha$, Zeeman field $B$ and chemical potential $\mu$) and is described by a single-band model $H_{L/R}=\frac{p_x^2}{2m}-\mu-\frac{\alpha}{\hbar}\sigma_y p_x+B\sigma_x$ (with $p_x=-i\hbar\partial_{x}$ the momentum operator and $\sigma_i$ Pauli matrices in spin space). \editE{These NWs undergo a topological phase transition at $B_c\equiv\sqrt{\Delta^2+\mu^2}$ with the appearance of MBSs at their edges, see blowup in Fig. \ref{fig:evenodd} (a).} $\Delta(x,\varphi)=i\sigma_y\Delta e^{\pm i\varphi/2}$ ({where the $\pm$ corresponds to $x\in L/R$, respectively}) is the induced pairing term \cite{Cayao:PRB17,Avila-accompanying}. 
While $V_J(\varphi)\sim -E_J \cos(\varphi)$ {is a good approximation at $B=0$} in the $T_N\rightarrow 0$ tunneling limit, it can strongly deviate from this form under relevant experimental conditions \cite{PhysRevA.69.062320,PhysRevB.97.060508,Avila-accompanying}.

Starting from Eq. \eqref{microscopic}, our goal is to derive a {quantitatively precise} low-energy {approximation for $V_J$ that retains both} standard Josephson events due to Cooper pair tunneling, as well as anomalous Majorana-mediated events where a single electron is transferred across the junction. While the \textit{total} fermion parity {$n\,\mathrm{mod}\, 2$ is conserved (where $n = n_L + n_R$ and $n_{L,R}$ are the fermion occupations in the left/right segments of the junction)}, an anomalous tunneling term changes the parity {$n_{L/R}\,\mathrm{mod}\,2$} on each superconducting left/right segment. 
For instance, assuming even global parity, Majorana-mediated tunnelling corresponds to the process $|0\rangle\equiv|n_L=0,\,n_R=0\rangle  \Longleftrightarrow|1\rangle\equiv|n_L=1,\,n_R=1\rangle $. (In what follows, even/odd parity will always refer to the partial fermion parity $n_L\,\mathrm{mod}\,2 = n_R\,\mathrm{mod}\,2$).  Physically, this {suggests that to compute an effective {low-energy $V_J(\varphi)$}, 
it is convenient to distinguish two contributions, $V_J(\varphi) = V_J^{bulk}(\varphi) + H_\textrm{BdG}^{sub}$}. \editRR{The first one takes into account the bulk contribution of BdG levels {\emph{above} the gap to the ground state (GS) energy 
$ V_J^{bulk}(\varphi)=-\sum_{\epsilon_p>\Delta}\epsilon_p(\varphi)$}.  The second contribution corresponds to the subgap sector (four subgap states resulting from two electron-hole copies of two spin-resolved subgap states), which can be described 
in terms of four Majorana operators, $\gamma_{1,2}\in L$ and  $\gamma_{3,4}\in R$, interacting pairwise \cite{PhysRevLett.108.257001,PhysRevB.86.140504}} 
\begin{eqnarray}
\label{subgap}
&H_\textrm{BdG}^{sub}(\varphi)=i\lambda_{12}\gamma_1\gamma_2+i\lambda_{13}(\varphi)\gamma_1\gamma_3%
+i\lambda_{14}(\varphi)\gamma_1\gamma_4\nonumber\\
&+i\lambda_{23}(\varphi)\gamma_2\gamma_3+%
i\lambda_{24}(\varphi)\gamma_2\gamma_4+i\lambda_{34}\gamma_3\gamma_4.%
\end{eqnarray}

\textit{Projection method--}
{To derive the different $\lambda_{ij}$ from microscopics, we first calculate the Nambu spinors $\check{\psi}_{is}^0$ of the full/empty ($i=\pm$) lowest subgap eigenstate of each decoupled $s=L,R$ segments. These states span a fermion basis $\{c_L,c_L^\dagger,c_R,c_R^\dagger\}$ of our 4-dimensional projection space, in terms of which we can write fermion numbers as} $n_L=c_L^\dagger c_L=(1+i\gamma_1\gamma_2)/2$ and $ n_R=c^\dagger_Rc_R=(1+i\gamma_3\gamma_4)/2$. Performing the low-energy projection
\begin{eqnarray}
	(\mathcal{H}^{-1})_{i's', is}&=&\langle \psi^0_{i's'}|G(\omega=0)|\psi^0_{is}\rangle,
\label{eq:lowenergy}
\end{eqnarray}
where $G(\omega)=(\omega+i\varepsilon-H_\textrm{BdG})^{-1}$ is the resolvent of the full BdG Hamiltonian, 
Eq.~(\ref{subgap}) can be written as $ H_\textrm{BdG}^{sub}=\frac{1}{2}\check{\psi}^{0\dagger}\mathcal{H}\check{\psi}^{0}$, {which yields the different $\lambda_{ij}$}. After projecting onto the parity basis {$\{|p\rangle\} = \{|0\rangle,|1\rangle\}$}, 
the final effective Hamiltonian reads
\begin{equation}
\label{Majorana-Transmon hamiltonian2}
H=[4E_C\left(-i\partial_{\varphi}-n_g\right)^2+V_J^{bulk}(\varphi)]\mathbb{1}+ V_J^{sub}(\varphi).
\end{equation}
It describes two different parity copies of a superconducting island, which are mixed through a non-diagonal term $V_{J\,p'p}^{sub}(\varphi) = \langle p'|H_\textrm{BdG}^{sub}|p\rangle$. The parity content of the eigenstates of Eq. (\ref{Majorana-Transmon hamiltonian2}) \editRR{is calculated} by a projection onto the parity axis defined by $\hat\tau_z\equiv|0\rangle\langle 0|-|1\rangle\langle 1|$. 

\editRR{Since Eq. (\ref{Majorana-Transmon hamiltonian2}) is derived by projecting $H_\textrm{BdG}$, the resulting coefficients in Eq. \eqref{subgap} depend on all the microscopic details of the problem, $\lambda_{ij}(B, L_S, \mu, \alpha)$,  instead of just being constant parameters like in \editEE{previous} effective models \cite{ginossar2014microwave,PhysRevB.92.075143}. Thus, $H$ is  able to describe, in particular, the full magnetic field evolution of the junction (from trivial to topological),} 
  \editRR{which is governed by three relevant $B$-field-dependent microscopic energy scales \cite{Avila-accompanying}: the Josephson coupling  $E_J = \int_0^{2\pi}\frac{\textrm{d}\varphi}{\pi}\left[ V^{bulk}_J(\varphi)+{V_{J\,{00}}^{sub}}(\varphi)\right]\,\cos(\varphi)$, the energy splitting between different fermionic parities $\delta=V_{J\,{11}}^{sub}(\varphi)-V_{J\,{00}}^{sub}(\varphi)|_{\varphi=0}$ and the Majorana coupling $E_M=\int_0^{2\pi}\frac{\textrm{d}\varphi}{\pi}\, V_{J\,{01}}^{sub}(\varphi)\cos(\varphi)$. Moreover, we can also analyze other possibilities, including the emergence of robust zero modes at $B\ll B_c$ (without a topological bulk) in wires with spatially inhomogeneous $\mu(x)$ \footnote{\editRR{The possibility of trivial Andreev levels has also been analyzed within the framework of the effective low-energy model in Ref. \cite{PhysRevB.100.241408}. Note, however, that the scenario discussed in \editEE{that reference}, a quantum dot in a topological insulator wire, cannot lead to the robust zero modes discussed here but only \editEE{to} accidental zero energy crossings of Andreev levels.}}}.

\emph{Results}--%
In what follows, the ratio $E_J/E_C$ and the plasma frequency $\omega_{pl}\equiv\sqrt{8E_JE_C}/\hbar$ {are defined respect to the zero-field junction, for which a microscopic calculation of $I_c$} \cite{San-Jose:NJP13,San-Jose:PRL14,Cayao:PRB17} gives $E_J\equiv\hbar I_c(B=0)/2e$ at fixed $E_C$. \editEE{In Fig. \ref{fig:evenodd}} the \editRR{calculated} fermionic parity content is represented by different colors (even/odd=blue/orange for \editE{$\langle\hat\tau_z\rangle=\pm 1$}). In the CPB limit 
each even (odd) parabola in the spectrum, \editEE{see Fig.~\ref{fig:evenodd}(b),} has a minimum at \editE{$n_g=m+n_g^0$}, where $m\in \mathbb{Z}$ and $n_g^0=0\ (0.5)$ (only the $m=0,1$ cases are shown). \editRR{Odd parabolas are energy-shifted from even ones by an amount $\delta$. For $B=0$ (not shown), the spectrum is $2e$-periodic and the energy shift is exactly $\delta=2\Delta$}, 
since 
each extra fermion must overcome an energy gap $\Delta$ on each NW segment. 
Increasing $B$, the gap gets reduced until an odd GS around $n_g=0.5$ is possible when $\delta<E_C$. In the $\delta\rightarrow 0$ limit (dotted curve), both parity sectors have minima at zero energy and the periodicity becomes $e$ \cite{Albrecht2016,Shen2018}. 
Furthermore, MBSs coherently mix parities around $n_g=0.25$ and $n_g=0.75$ \cite{ginossar2014microwave,PhysRevB.92.075143,10.21468/SciPostPhys.7.4.050}. 
\editRR{In the transmon limit with $\delta=0$, Fig.~\ref{fig:evenodd}(c), strong parity mixing occurs for all $n_g$, which manifests as splitting of the lines with almost zero parity polarization $\langle\hat\tau_z\rangle\approx 0$. At a different $B$ corresponding to 
$\delta\neq 0$, Fig.~\ref{fig:evenodd}(d), the transmon spectrum shows charge dispersion and an overall $2e$-periodicity. This result illustrates how nominally identical transmons depend on microscopic properties of the NWs (in this case, 
the ratio $\xi_M/L_S$, with $\xi_M$  the Majorana coherence length \cite{Mishmash:PRB16}, which governs $\delta(B)$ \footnote{\editRR{This claim is only true for spatially-homogeneous wires since a smooth $\mu(x)$ can lead to $\delta\to 0$ irrespective of the ratio $\xi_M/L_S$, as we discuss in Fig. 3.}}). Moreover, the finite $\delta\neq 0$ polarizes the GS parity (now of odd character for all $n_g$).}

\edit{Next, we focus on the magnetic field evolution of the island spectrum}. 
In the CPB regime, this evolution strongly depends on gate (owing to {the large} charge dispersion). Since parity mixing occurs near  $n_g=0.25$, the $B$-field dependence at this gate considerably differs from the one at $n_g=0.5$, where parity mixing is negligible, compare Fig.~\ref{fig:evenodd}(e) and Fig.~\ref{fig:evenodd}(f). \editEE{On the other hand,} \editRR{in the transmon limit the spectrum mimics Majorana oscillations. After each parity crossing, see arrows in Figs.~\ref{fig:evenodd}(g,h), the parity polarization $\tau_z$ jumps, Figs.~\ref{fig:evenodd}(i,j), which reflects changes in the parity character of the GS (change of color from blue to orange and back). This behavior occurs for all $n_g$. Switches of GS parity are possible since, for the single band case considered here, $E_M\sim E_J$ \cite{San-Jose:NJP13,San-Jose:PRL14,Cayao:PRB17,Tiira:NC17,Avila-accompanying}. Consequently, in a transmon regime with $E_J/E_C\gg1$, the ratio $E_M/E_C$ is not small \footnote{\editRR{Considering typical critical current values $I_c\sim 0.2 I_0$, with $I_0=e\Delta/\hbar$ the maximum supercurrent of a single open channel, this gives $E_J\sim 0.1\Delta\sim 25\mu$eV$\sim 6$GHz, assuming an induced gap of the order of $\Delta\sim 250\mu$eV. Since $E_M\sim E_J$ for a topological single channel,  and considering typical charging energies for transmons $E_C\sim 200-300$MHz \cite{PhysRevLett.115.127001,PhysRevLett.115.127002,PhysRevB.97.060508}, the junction is naturally in the $E_M/E_C\gg1$ regime}}}. 
 \editRR{This relevant regime has hitherto remained unexplored, even at the level of the effective models in Refs. \cite{ginossar2014microwave,PhysRevB.92.075143,10.21468/SciPostPhys.7.4.050}, which focus on the opposite $E_M/E_C\ll1$ regime where the large charging energy prevents changes in the GS parity \cite{Avila-accompanying}. This would require much larger $E_C$ (which is detrimental since it induces charge dispersion) or very small Majorana couplings $E_M\ll E_J$. The latter can be, however, somewhat difficult to reach in a few-channel topological wire: while more than one channel, but not many, can contribute to $E_J$ \cite{PhysRevB.97.060508,Goffman_2017,PhysRevX.9.011010}, the value of $E_M$ is of the order of the topological minigap \cite{Avila-accompanying}. These parity switches in the GS have important consequences for MW spectroscopy, as we discuss next. } 
\begin{figure}
\includegraphics[width=\columnwidth]{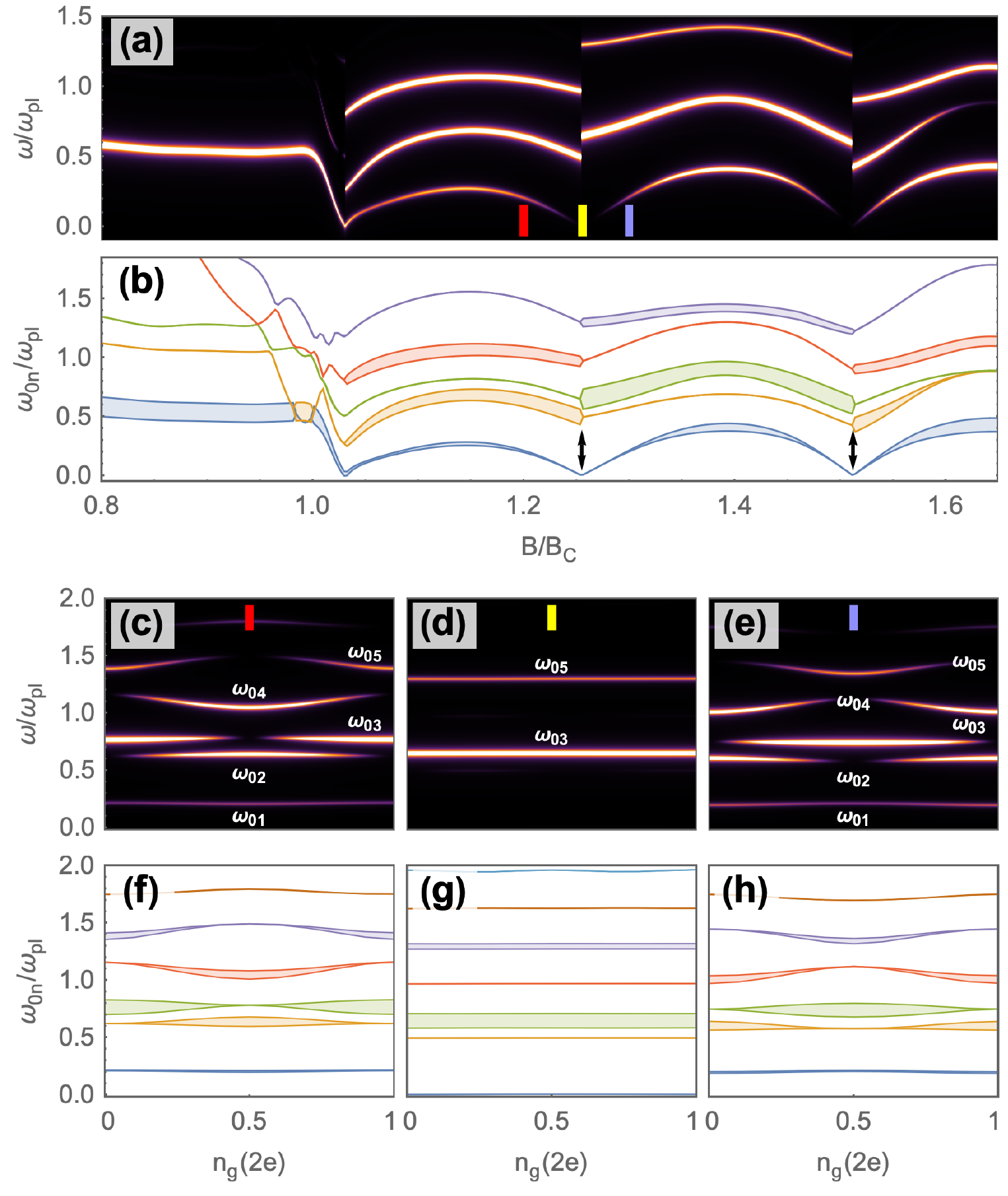}
\caption{\label{fig:MW} \textbf{MW spectroscopy of a NW-based transmon}. (a) Contour plot of \editE{MW absorption spectrum} $S_N(\omega)$ versus $\omega$ and $B/B_c$ at $n_g=0.5$. Bright lines signal allowed transitions in the MW response. {Spectral holes in the $n=1$ transition line and abrupt jumps in the $n>1$ ones}, where lines suddenly disappear/appear, coincide with parity crossings in the GS owing to Majorana oscillations. (b) Transition frequencies and spectral weights (shadowed widths). At minima of the oscillations (arrows), the spectral weight of different transitions gets exchanged. (c-h) $n_g$ dependence. Same parameters as the transmon in Fig. \ref{fig:evenodd}.}
\end{figure}
\begin{figure}
\includegraphics[width=\columnwidth]{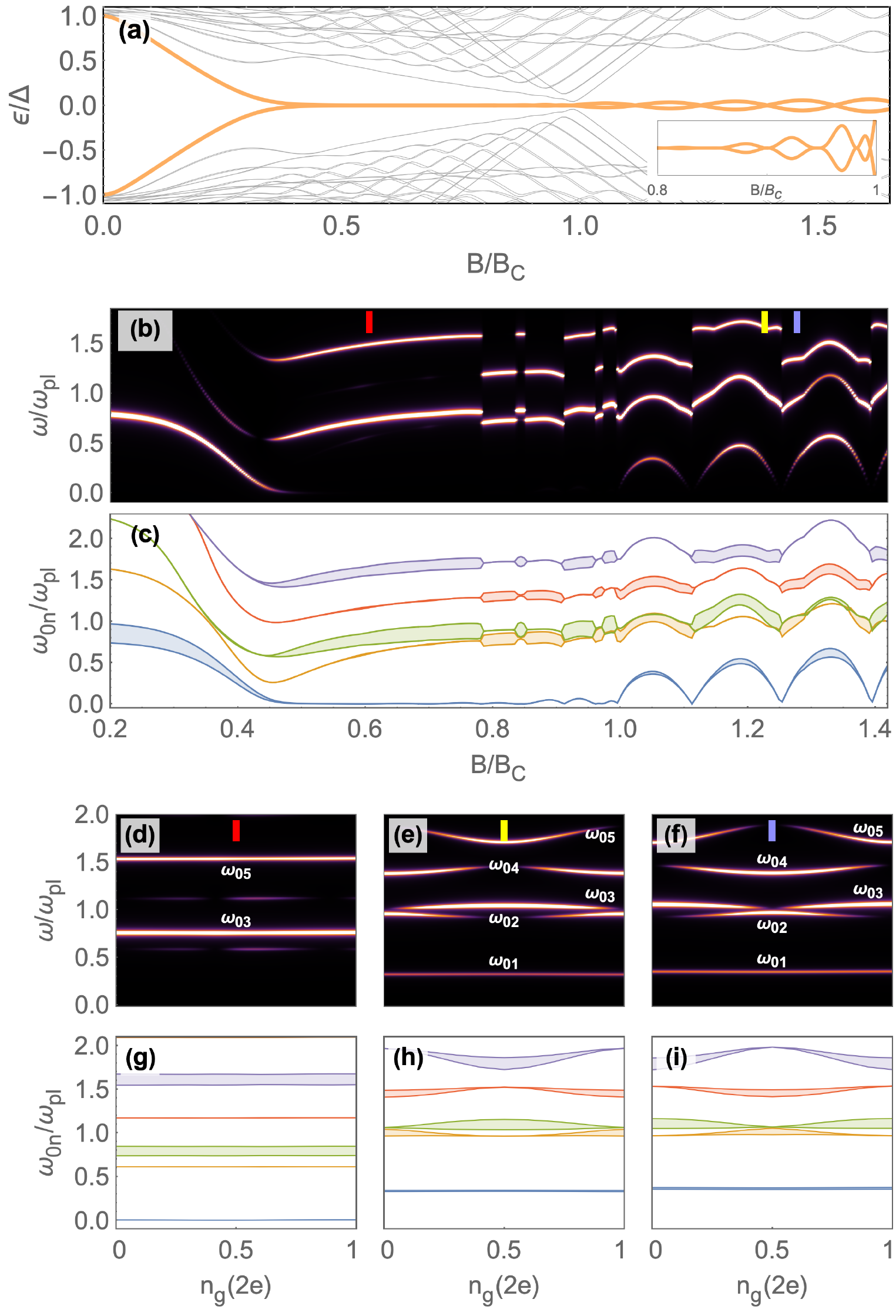}
\caption{\label{fig:smooth}  \editRR{ \textbf{MW spectroscopy of a junction containing non-topological robust zero modes.} (a) The BdG spectrum of a NW with \editEE{a smooth} $\mu(x)$ (which interpolates between $\mu=0.5$ meV and $\mu=2$ meV at each NW segment) shows robust zero modes for $B<B_c$. Inset: blow up near $B\lesssim B_c$ where the zero mode starts to develop oscillations. (b,c) The $B$-field dependence of the MW response (shown here at $n_g=0.5$) of a transmon containing such wires is distinct depending on whether it is governed by \editEE{the trivial mode pinned to zero at $B\lesssim 0.8B_c$ or the oscillating Majoranas} at $B>B_c$. While the former leads to an overall standard transmon response (d,g), the later shows parity switches, (e,h) and  (f,i), as discussed in Fig. \ref{fig:MW}. Same parameters as Fig. \ref{fig:MW}.}}
\end{figure}
\emph{MW spectroscopy}-- 
The linear-response MW spectrum can be written as $S_N(\omega)=\sum_n |\langle n|\hat N|0\rangle|^2 \delta(\omega - \omega_{n0})$, where $\Psi_n$ are the eigenstates of Eq. \eqref{Majorana-Transmon hamiltonian2} and $\omega_{0n}=(E_n-E_0)/\hbar$. \edit{In what follows we focus on the transmon limit with $E_M/E_C \gtrsim 1$. 
A full discussion about other relevant regimes, including the CPB and the transmon regime with {$E_M/E_C \ll 1$}, can be found in Ref. \cite{Avila-accompanying}.}
In  Fig.~\ref{fig:MW}(a) we plot the magnetic field dependence of the MW spectrum of a transmon. 
Majorana oscillations and parity switches in Figs.~\ref{fig:evenodd}(h,j) are found to result in abrupt spectroscopic changes  \edit{at $B$-fields where Majorana oscillations have nodes. They include spectral \emph{holes} in the first transition $\omega_{01}$ \editRR{(a transition within the ground state manifold which is visible thanks to the Majorana-induced parity mixing)} accompanied by higher transition lines suddenly disappearing/appearing. The spectral holes can be understood as exact cancellations of spectral weight owing to \editRR{parity degeneracy at crossings, see Fig.~\ref{fig:evenodd}(j)}. This is illustrated in Fig.~\ref{fig:MW}(b) where we plot the transition frequencies weighted by their respective matrix element (represented as the width of the line). Together with the absence of spectral weight of the $\omega_{01}$ transition at crossings, there is a complete spectral weight transfer between higher energy transitions, where one thick line becomes thin or viceversa (see arrows). 
Figs.~\ref{fig:MW}(c-e) show the $n_g$ dependence for three fixed $B$-fields [colored bars in (a)] across a parity crossing. Interestingly, all the spectroscopic features before and after the crossing [(c) and (e), respectively] \emph{are shifted exactly by one $e$ unit}, which reflects the change of parity of the GS. See also the transitions weighted by their matrix elements in Figs. \ref{fig:MW}(f-h). This results in distinct spectroscopic features like spectral holes that shift from half-integer to integer values of $n_g$ and viceversa, and changes of curvatures of the involved transitions. Right at the $\delta=0$ parity crossing, Figs. \ref{fig:MW}(d,g), we recover a standard transmon spectrum. Such unique behavior of transmon spectra across a Majorana oscillation should provide a strong signature of the presence of MBSs and their associated parity crossings.}

\editRR{The above behavior has to be contrasted with the spectra resulting from \editEE{non-oscillating} zero modes. Figure \ref{fig:smooth}(a) shows a typical BdG spectrum where a spatially smooth potential $\mu(x)$ in the NW results in a robust zero mode \editEE{starting at} $B\ll B_c$. These non-topological  zero modes are ubiquitous in experiments and have spurred a longstanding `trivial versus topological" debate in the literature regarding tunneling spectroscopy measurements of zero bias anomalies \cite{Moore:PRB18,Penaranda:PRB18,Vuik:SP19,Avila:CP19,prada2019}. Interestingly, the MW response of such robust $B<B_c$ zero modes is very distinct from the one we discussed previously for oscillating Majoranas. Here, the overall response of the $B<B_c$ zero mode is that of a transmon (but with a main line $\omega_{03}$ corresponding to a parity-conserving transition from an \emph {odd} GS) and a largely suppressed parity-mixing line $\omega_{01}$, see Figs. \ref{fig:smooth}(b,c). At a fixed $B<B_c$ field (red bar), the $n_g$ dependence is that of a transmon with no charge dispersion, see Figs. \ref{fig:smooth}(d,g). In contrast, the appearance of parity crossings which fully develop into oscillating Majoranas at larger $B$-fields [e. g. blue and yellow bars in Fig. \ref{fig:smooth}(b)] results again in an oscillating  $\omega_{01}$ line, switches in the higher transitions and one-$e$ shifts in the gate response, Figs. \ref{fig:smooth}(e,h) and Figs. \ref{fig:smooth}(f,i). Both the amplitude of the oscillations and the intensity of the lowest intraband transition $\omega_{01}$ grow with $B$-field, providing yet another clear signature as compared to the low frequency response resulting from trivial zero modes at $B<B_c$.}

\editRR{In summary, our results demonstrate that MW spectroscopy of NW-based transmon qubits is a powerful tool to detect the presence of \editEE{oscillating} MBSs in the JJ junction. Repeated parity switches in the MW spectrum against magnetic field reveal unambiguously the existence of a subgap state oscillating around zero, which could aid in distinguishing topological Majoranas from other non-oscillating types of zero modes \cite{prada2019}, thus providing an alternative to tunneling spectroscopy.
Our projection method can be readily extended to other relevant regimes \cite{Avila-accompanying}}. Other geometries currently under intensive experimental study, including \editEE{split junctions \cite{Avila-accompanying},} junctions with quantum dots \cite{bargerbos2019,kringhj2019} and superconducting islands in the fluxonium regime \cite{pitavidal2019}, should be the subject of future studies. 
\acknowledgements
We thank Bernard van Heck for his input. Research supported by the Spanish Ministry of Science and Innovation through Grants PGC2018-097018-B-I00, FIS2016-80434-P (AEI/FEDER, EU), BES-2016-078122 (FPI programme), and RYC-2011-09345 (Ram\'on y Cajal programme). The EU Horizon 2020 research and innovation programme (FETOPEN Grant Agreement No. 828948) and the CSIC Research Platform on Quantum Technologies PTI-001 are also acknowledged.

\bibliography{biblio}

\end{document}